\documentclass[onecolumn]{svjour3}     
\pdfoutput=1
\smartqed  
\usepackage{graphicx}
 \usepackage{mathptmx}      
\usepackage{latexsym}
\usepackage{amsfonts}
\newcommand{\bea}{\begin{eqnarray}}
\newcommand{\eea}{\end{eqnarray}}
\newcommand{\beq}{\begin{equation}}
\newcommand{\eeq}{\end{equation}}
\newcommand{\nn}{\nonumber}
\newcommand{\eqref}[1]{(\ref{#1})}
\newcommand{\C}[1]{{\mathcal{#1}}}

\newcommand{\abs}[1]{\vert #1\vert}

\newcommand{\expect}[2]{\left\langle \, #1\, \right\rangle_{#2}}
\newcommand{\CDT}{{\mathrm{CT}}}

\newcommand{\RR}{{\mathrm{R}}}
\newcommand{\GRT}{{\mathrm{T}}}
\newcommand{\RRP}{{\mathrm{R'}}}
\newcommand{\CGW}{{\mathrm{GW}}}

\newcommand{\half}{\frac{1}{2}}
\newcommand{\quarter}{\frac{1}{4}}

\newcommand{\braket}[2]{\langle #1\vert\, #2\rangle}

\newcommand{\event}[2]{{\C #1}_{#2}}

 \journalname{Journal of Statistical Physics}
\begin{document}

\title{On the spectral dimension of causal triangulations}




\author{Bergfinnur Durhuus       \and Thordur Jonsson \and \\ John F Wheater
}


\institute{B. Durhuus \at
              Department of Mathematical Sciences, University of Copenhagen\\ Universitetsparken 5, 2100 Copenhagen \O, Denmark\\
              \email{durhuus@math.ku.dk}           
           \and
           T. Jonsson \at University of Iceland, Dunhaga 3, 107 Reykjavik, Iceland\\
              \email{thjons@raunvis.hi.is}
              \and
              J.F. Wheater\at Rudolf Peierls Centre for Theoretical Physics, University of Oxford, 1 Keble Road, Oxford OX1 3NP, UK\\ Tel.: +44 1865 273961, Fax.: +44 1865 276667\\ \email{j.wheater@physics.ox.ac.uk}
}

\date{Received: date / Accepted: date}

\maketitle

\begin{abstract} We introduce an ensemble of infinite causal
  triangulations, called the uniform infinite causal triangulation,
  and show that it is equivalent to an ensemble of infinite 
  trees, the uniform infinite planar tree. It is proved that in both
  cases the Hausdorff dimension almost surely equals $2$. The infinite
  causal triangulations are shown to be almost surely recurrent or,
  equivalently, their spectral dimension is almost surely less than or
  equal to $2$. We also establish that for certain reduced versions of the
  infinite causal triangulations the spectral dimension
  equals $2$ both for the ensemble average and almost surely. The triangulation ensemble
  we consider is equivalent to the causal dynamical triangulation model of
  two-dimensional quantum gravity and therefore our results apply  to that model.

\end{abstract}

\section{Introduction}
\label{intro}

The behaviour of random walks, or equivalently diffusion,  on random graphs 
has been studied intensively in recent times.  
The motivation for doing so has come from many different areas of physics.  For example 
these problems play a central role in 
the study of random media and have been investigated both by numerical
and analytic methods  \cite{benAv:2000}.  In this paper we are
concerned with the ensembles of random graphs which arise in
discretized quantum gravity models (see \cite{Ambjorn:1997di} for an
introduction) and we will establish some exact results for
two-dimensional versions of these models which are sufficiently
tractable.  These  are of interest in their own right but of course
one might hope that they can also  provide insight into the higher
dimensional models.

The connection between theories of gravity and ensembles of random graphs is made through the metric tensor $g_{\mu\nu}$ which is the dynamical degree of freedom. In classical general relativity
 the metric satisfies Einstein's equations and for any given set of consistent 
 initial conditions there is a unique evolution of the metric in time. 
 Quantization using the path integral formalism then amounts, at least naively, 
to forming a quantum amplitude describing the evolution of the metric 
 from $g^A_{\mu\nu}$ at $t=0$ to $g^B_{\mu\nu}$ at $t$ with amplitude
\beq \braket{g_B,t}{g_A,t=0}= \sum_{g\in\Gamma} \exp(i S[g]/\hbar),\label{qgA}\eeq
where $\Gamma$ is the set of all possible metrics satisfying $g=g_A$
at $t=0$ and   $g=g_B$ at $t$, and $S[g]$ is the action (the natural
choice for which is the Einstein--Hilbert action which in two dimensions
with fixed topology consists of the cosmological constant term alone). Note that we have
made a number of assumptions here concerning the consistent definition
of $t$. To evaluate the amplitude (\ref{qgA}) requires a systematic
way of describing the set $\Gamma$. The discretized random surface is
one way of doing this \cite{Ambjorn:1997di}. For simplicity consider a
two-dimensional manifold with euclidean metric (so this is not really
gravity which should have a lorentzian metric) and spherical
topology. Such a manifold can be triangulated with $N\ge 2$ triangles;
the idea is that by taking $N\to\infty$ in an appropriate way we can
recover a continuum space. The  metric is associated with the
triangulation by  supposing that all triangles are equilateral of side
$a$ and defining  the geodesic distance between any two points as $La$
where $L$ is the number of edges in the shortest path connecting
them. Then every distinct triangulation $T$ leads to a distinct metric
and the vacuum amplitude is given by 
\beq Z= \sum_{T\in\C P} \exp(-S_T)\eeq
where we have  set $\hbar=1$, $S_T$ is the discretized equivalent of the continuum action,
  and $\cal P$ is the set of all distinct
triangulations of the sphere -- or equivalently the planar random
graphs with all  vertices having degree 3.   
Many objects of interest have been calculated in this particular model
which is often known as `two-dimensional euclidean quantum gravity';
it has a scaling limit in which $a\to 0$ and $N\to\infty$ in such a
way that  a non-trivial continuum model results and we refer the
reader to \cite{Ambjorn:1997di} for details. However there are
problems with this model as a theory of gravity some of which seem to
arise as a consequence of the absence of any notion of causality in
the theory. 
The Causal Dynamical Triangulation (CDT) model was invented 
\cite{Ambjorn:1998xu} to build in causality  from the start by imposing a well defined
temporal structure. This is done by restricting $\cal P$ to  random
triangulations  which can be consistently sliced perpendicular to one
direction (the time-like direction) and in which topology change is
forbidden for sub-graphs lying in the other (spacelike) direction --
these graphs are fully defined in Section \ref{subsec:CDT} below. 
 The idea can be applied to space-times of two or more dimensions;
 unfortunately it becomes progressively more difficult with increasing
 dimension to obtain analytic results  although  much has been learned
 by doing numerical simulations \cite{Ambjorn:2005db,Ambjorn:2005qt}.

The geometry  of the ensembles of graphs appearing in these gravity models can 
be characterized in part by universal quantities of which 
 the most basic  is the dimensionality.  There are different notions
 of dimension. 
The simplest one to evaluate is usually  the Hausdorff dimension
\(d_h\) of a graph $G$, 
which is defined provided   the volume $V_G(R)$ enclosed within a ball of a radius \(R\) takes the form
\begin{equation}\label{Hausdorffdim}
 V_G(R)  \sim R^{d_h}
\end{equation}
at large $R$.
  The spectral dimension is defined to be \(d_s\) provided the
probability $p_G(t)$ that a random walker on a graph $G$ returns to the point of origin after a time \(t\) takes the form 
\begin{equation}\label{spectraldim}
 p_G(t)  \sim t^{-d_s/2}
\end{equation}
at large time. 
For the fractal geometries we are interested in it is not necessarily
true that all definitions of dimension agree.  
The spectral dimension probes different aspects of the long range properties of 
graphs from the Hausdorff dimension; clearly it is in some sense a measure of how 
easy it is for a walker to travel between different regions of the graph rather 
than a static measure of how large those regions are.  It is important to note that the
definitions \eqref{Hausdorffdim} and \eqref{spectraldim} only make
sense for infinite (connected) graphs. This is obvious for
\eqref{Hausdorffdim} while for \eqref{spectraldim} it is easy to see
that $p_G(t)$ tends to a non-vanishing constant for $t\to\infty$ if
$G$ is finite.  

For fixed graphs which 
satisfy certain uniformity conditions  it is known that
\beq d_h\ge d_s\ge\frac{2d_h}{1+d_h}\eeq
provided both dimensions exist, see for example \cite{Coulhon:2000}.Those uniformity
conditions are not necessarily applicable to random graphs although  this relation is 
satisfied in at least some examples of ensemble averages of random geometries 
\cite{Durhuus:2005fq}. For random graphs in general the dimensions $d_h$ and
$d_s$ can be defined either by replacing the left hand sides of
\eqref{Hausdorffdim} and \eqref{spectraldim} by their ensemble averages or,
 more ambitiously, by  establishing that individual graphs
almost surely have a definite value of $d_h$ or $d_s$. We shall focus mainly 
on the latter point of view in this paper. The methods 
we employ build on those used in earlier studies of random walk on random combs 
\cite{Durhuus:2005fq} and on generic random trees
\cite{Durhuus:2006zz,Durhuus:2006vk}. Related results on the recurrence of random planar graphs with  bounded  vertex degree have been
obtained in \cite{EJP2001-23}. In this paper  we consider  graphs that do not have bounded vertex
degree, although they do have other special characteristics, and so in some sense extend these results.


This paper is organized as follows. In Section \ref{sec:ensembles} the ensembles 
of graphs that we consider in this paper are 
introduced and the relationships between them and tree ensembles constructed 
from Galton Watson processes explained.  Section \ref{sec:hausdorff} discusses 
the Hausdorff dimension of these ensembles while in Section \ref{sec:recurrence} 
it is proved that the two-dimensional causal dynamical triangulation ensemble 
is recurrent and therefore that its spectral dimension is bounded above by 2.   
In Section \ref{sec:spectral} we prove that the spectral dimension in 
 the related radially reduced model is exactly 2.
In the final section we discuss the significance of our results.    

\section{Ensembles of random graphs}
\label{sec:ensembles}
A random graph $(\C G,\mu)$ is a set of graphs $\C G$ equipped with a probability measure
$\mu$. In the following we assume the graphs in $\cal G$ to have a marked
vertex called the \emph{root}. We shall discuss several measures $\mu$
or $\mu_X$ and will denote the corresponding expectation 
by $\expect{\cdot}{\mu}$ or $\expect{\cdot}{X}$.
 The ensembles that we consider are all related to the generic random tree 
$(\C T,\mu_\infty)$ which was studied in \cite{Durhuus:2006vk} and which we first review.

\subsection{The generic random tree}
\label{sec:2.1}

A rooted tree $T$ is a connected planar graph consisting of vertices $v$ of finite degree connected by edges but 
containing no loops; the root $r$ is a special marked vertex connected to only one edge and the smallest 
rooted tree consists of the root and one other vertex. We denote the number of edges in a tree by $\abs T$. The 
set of all trees $\C T$ contains the set of finite trees ${\C T}_f =\bigcup_{N\in {\mathbb N}} {\C T}_N$ 
where ${\C T}_N=\{ T\in  {\C T} : {\abs T} =N\} $ 
and the set of infinite trees $\C T_\infty$. 

A Galton Watson (GW) process is defined by offspring probabilities which are a sequence of non-negative numbers 
$p_0\ne 0, p_1, p_2,\ldots$, with $p_i >0$ for at least one $i\ge 2$. They are
conveniently encoded  in the generating function 
\beq f(x)=\sum_{n=0}^\infty   p_n x^n 
\eeq
which satisfies $f(1)=1$. 
We call the process \emph{critical} if $f'(1)=1$ and \emph{generic} if $f(x)$ is analytic in a neighbourhood of the unit disk.
Assigning the  probability $p_{\sigma_v-1}$ to the event that any  vertex $v\ne r$ has degree $\sigma_v$ a critical GW process 
induces a probability distribution $\mu_\CGW$ on $\C T_f$,
\beq 
\mu_\CGW(T ) = \prod_{v\in T\setminus r} p_{\sigma_v-1}.\label{treemeasure}
\eeq
We will call the ensemble $(\C T_f,\mu_\CGW)$ a critical Galton Watson ($\CGW$) tree.

\begin{figure}
\begin{center}
 \includegraphics[scale=0.5,angle=-90]{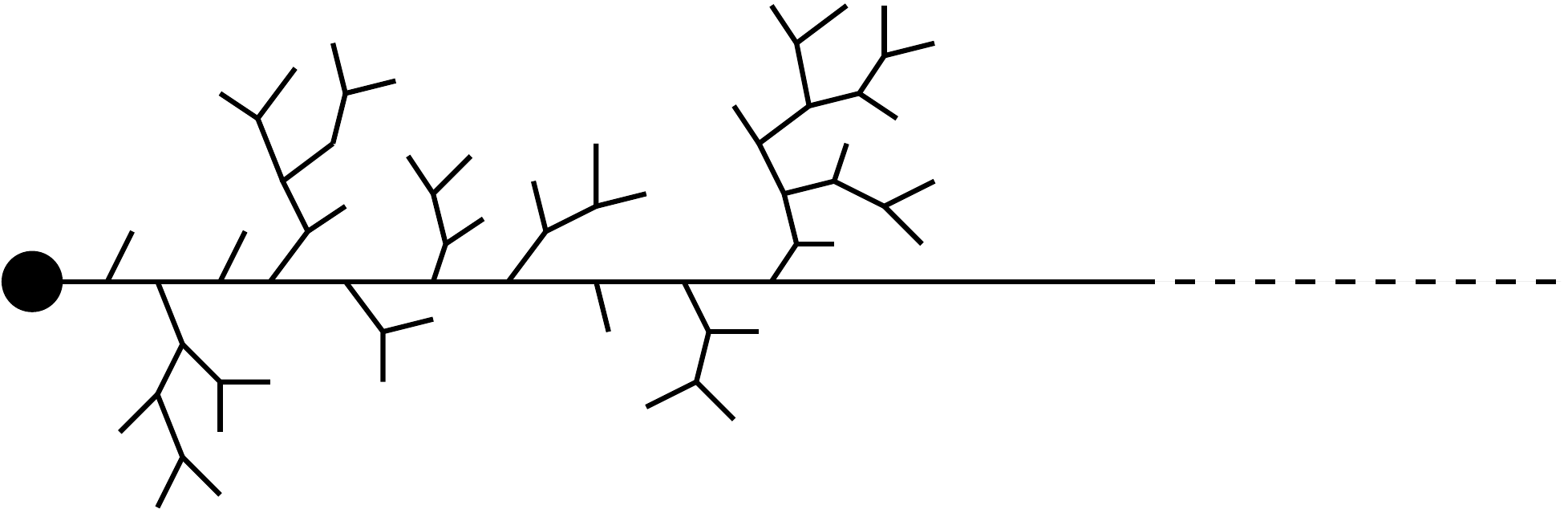}
 \end{center}
\caption{Example of $T\in{\C S}$ }
\label{fig:S}       
\end{figure}

Next we define the probability distribution $\mu_N$ on $\C T_N$ by
\beq \label{muNdef}
\mu_N(T)=Z_N^{-1}\prod_{v\in T\setminus r} p_{\sigma_v-1}
\eeq
where
\beq 
Z_N=\sum_{T\in\C T_N}\prod_{v\in T\setminus r} p_{\sigma_v-1}.
\eeq
We define the {\it single spine trees} to be the  subset $\C S$ of the infinite trees whose members consist of a 
single infinite linear chain $r, s_1,s_2,\dots$, called the \emph{spine}, to each vertex of which are attached a 
finite number of finite trees by identifying their root with that vertex. 
An example of a single spine tree is illustrated in Fig.{\ref{fig:S}}.
The following result was established in \cite{Durhuus:2006vk}. 
\begin{theorem}\label{thm1}
Assume that $\mu_N$ is defined as above as a probability measure on $\C T$ where $\{ p_n\}$ defines a
generic and critical GW process.  Then
\beq
 \mu_N\to\mu_\infty\quad as \quad N\to\infty\label{mudef}
 \eeq
where $\mu_\infty$ is a probability measure on $\C T$ concentrated on the set of single spine trees $\C S$.
The generating function for the probabilities for the number of finite
branches at a vertex on the spine is $f'(x)$. Moreover, the individual
branches are independently and identically distributed according to the original critical
GW process. 
\end{theorem}
The generic random tree associated to the given GW process is by
definition $(\C S , \mu_\infty)$.

In this context convergence of measures means that integrals of continuous
bounded functions on $\cal T$ converge, where continuity 
refers to a metric $d_{\cal T}$ according to which two trees $T$ and
$T'$ are close if they coincide on sufficiently large balls centred at the
root. More precisely one can use
\beq
d_{\cal T}(T,T') = \inf\{\frac{1}{R}\,:\; B_R(T)=B_R(T')\} \,,
\eeq
where the ball $B_R(G)$ of radius $R$ centred at the root of a graph $G$ is the subgraph
of $G$ spanned by the vertices at graph distance at most $R$ from
the root. The graph distance between two vertices in $G$ is as usual
the minimum number of edges in a path connecting them.  

Of particular interest in the following is the so-called \emph{uniform
  infinite planar tree} \cite{ISI:000186418300011} corresponding to
offspring probabilities 
\beq\label{geomp} 
p_n= 2^{-(n+1)}\,,\quad n\geq 0
\eeq
which are easily seen to satisfy the requirements for a generic, critical GW tree.
For this  random tree we shall use the notation $\bar\mu_N$ for
$\mu_N$ and $\bar\mu$ for $\mu_\infty$.

 We define the height $h(v)$ of a vertex $v$ in a graph
$G$ as the graph distance from $v$ to the root; the height  
$h(\ell )$ of an edge $\ell$ in $G$ as the minimum 
height of an end of $\ell$; and the height of a finite graph $G$ by 
\beq 
h(G)=\max_{ v\in G} h(v).
\eeq
 Given a tree $T$,  $D_k(T)$ is the set of vertices at height $k$ (so
 that $D_0=r$ and $D_1$ consists of the unique vertex which is the
 neighbour of the root); the number of vertices in $D_k(T)$  is
 denoted by $\abs{D_k(T)}$, whereas $\abs{B_R(T)}$ denotes the number
 of edges in $B_R(T)$. There are  a number of useful properties of $\mu_\CGW$ and $\mu_\infty$
which follow:

\begin{lemma}\label{height}
For large $R$
\beq 
\mu_\CGW(\{T\in \C T_f : h(T)>R\})=\frac{2}{f''(1)R}+O(R^{-2})
\eeq

\end{lemma}
\begin{proof} This is well known and the proof is given in e.g. \cite{harris:2002}.\end{proof}

\begin{lemma} \label{Dinv}There exists a constant $c>0$ such that 
\beq 
\expect {\abs{D_k}^{-1}}{\infty} \le \frac{c}{k}.
\eeq
Furthermore,
\bea 
\expect {\abs{D_k}}{\infty} &=& (k-1)f''(1)+1,\quad k\ge 1\label{Done}\\
\expect{\abs{B_k}}{\CGW}&=&k,\quad k\ge 1\label{BGW}\\
\expect{\abs{B_k}}{\infty}&=&\half k(k-1)f''(1)+k,\quad k\ge 1\label{BT}
\eea
\end{lemma}
\begin{proof}The proof is given in \cite{Durhuus:2006vk}, Proof of
    Lemma 5 and Appendix 2.\end{proof}

\subsection{Causal triangulations}
\label{subsec:CDT}
\begin{figure}
\begin{center}
 \includegraphics[scale=0.5,angle=-00]{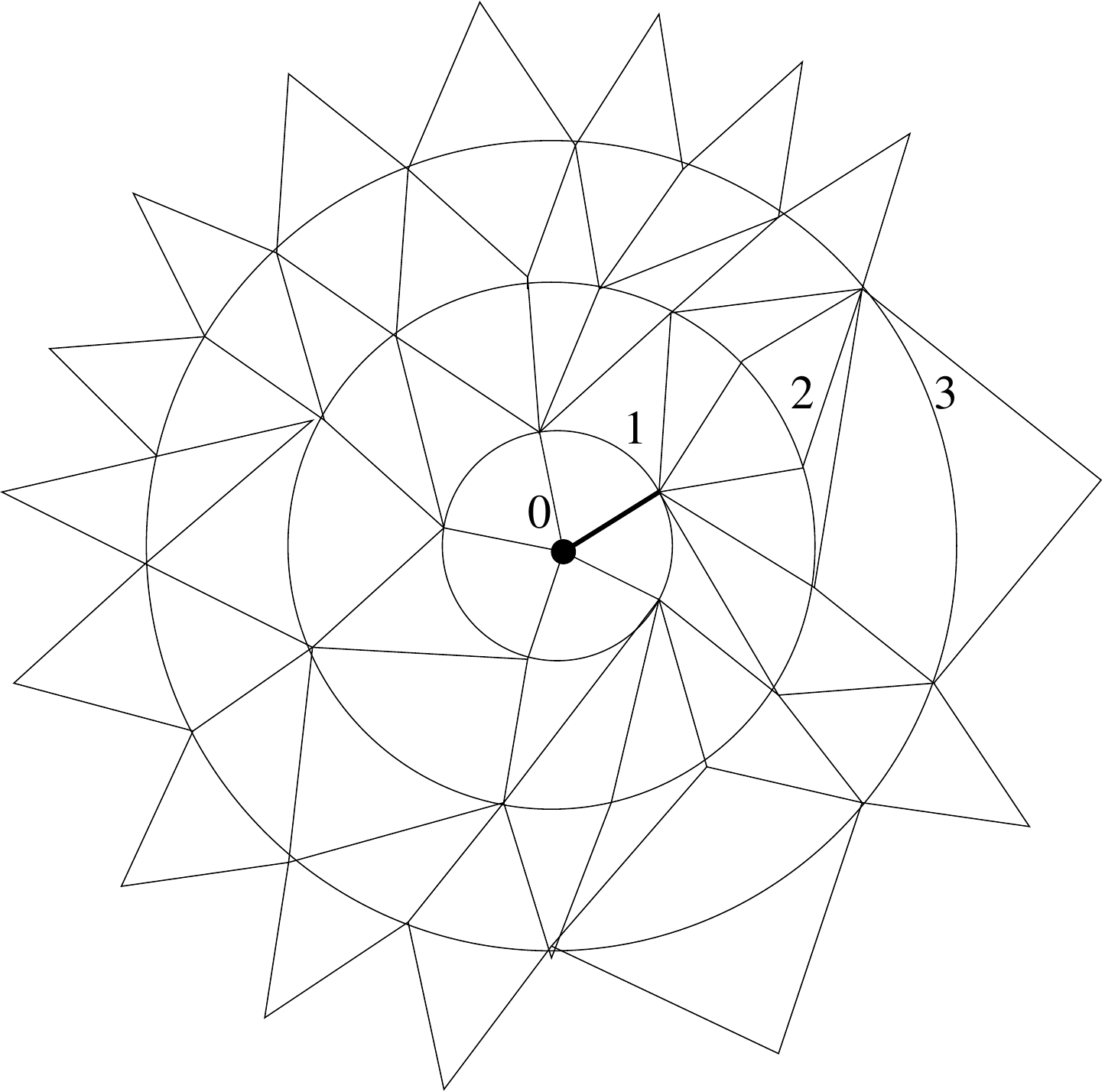}
 \end{center}
\caption{Example of $G\in{\C C_4}$; the numerical labels show the heights of the cycles and the root and marked edge are shown in bold. }
\label{fig:0}       
\end{figure}

In this sub-section we define the notion of a causal triangulation
(CT) and recall the definition of the model of causal dynamical
triangulations (CDT)  
introduced in \cite{Ambjorn:1998xu}.   
We say that a graph on $n$ vertices $v_1,v_2,\ldots ,v_n$ is a cycle if the edge set 
is 
\beq
\{ (v_1,v_2), (v_2,v_3),\ldots , (v_{n-1},v_n), (v_n,v_1)\}.
\eeq
We include the degenerate case
$n=1$ in which case we allow a loop so the unique edge is $(v_1,v_1)$.
Let $G$ be a rooted planar triangulation (i.e.\ a planar graph such
that all the faces, except 
possibly one, are triangles).  Let $S_0$ be the root vertex of $G$ and
$S_k$ the set of vertices at graph distance $k$ from the root,
$k=1,2,\ldots$.  We say that $G$ is 
a {\it causal triangulation} if  $S_k$ together with the edges in $G$
which join vertices in $S_k$, form a cycle for  $k<h(G)$ and, in the case $h(G)<\infty$, if the
cycle at height $h(G)-1$ is decorated by attaching to each edge a
triangle whose other two edges are not shared with any other triangle and whose 
vertex of order 2 belongs to the infinite face of $G$. The decoration of the highest cycle
with triangles is not essential to the definition of CTs but it is convenient when we come to 
consider the measure assigned to the graphs.
We denote by $\C C$ the collection of all causal triangulations, 
$\C C_K$ the elements in $\C C$ of height $K$, $\C C_f$ the collection
of all triangulations in $\C C$ of finite height and $\C C_\infty =\C
C\setminus \C C_f$. 
Note that any triangulation in $\C C_K$ is a triangulation of the
closed disk whose boundary vertices alternate in height between $K$
and $K-1$.  In this case  
the exterior face is not  a triangle.  The elements of $\C C_\infty$
can be viewed as triangulations of the plane with the property that
all the vertices 
at a fixed graph distance from the root form a cycle. For technical
reasons that will become clear below we will assume that one of the
edges emerging from the root vertex is marked and called the
\emph{root edge}. In particular, this eliminates accidental symmetries under rotations
around the root vertex. 
 An example of $G\in{\C C}_4$ is shown in  Fig.\ref{fig:0}. 
 
 Given a causal triangulation $G$ and $k<h(G)-1$ we will let $\Sigma_k$ denote the
 subgraph of $G$ which consists of  $S_k$  and $S_{k+1}$ together with
 the edges joining them. 
 Note that $\Sigma_k$ is a triangulation of an annulus. Furthermore,
 we denote the number of triangles in $G$  by $\Delta(G)$ and call it
 the \emph{area} of $G$. Note that 
 \beq \label{sigma} \Delta(\Sigma_k) =\abs{S_k}+\abs{S_{k+1}}\eeq
 (where $\abs{S_k}$ is the
   number of edges in $S_k$ and $\abs{S_0}=0$ by definition) and that the total area of $G\in{\cal C}_f$ is even
   and equals 
\beq\label{area}
\Delta(G) = 2\,\sum_{k=1}^{h(G)-1}\abs{S_k(G)}\;.
\eeq

In the CDT model \cite{Ambjorn:1998xu} each graph $G\in{\cal C}_f$ is assigned a weight 
\beq w(G)=  g^{1+\Delta(G)},\label{weight}\eeq
where $g$ is the fugacity for triangles, 
and the grand canonical partition function is
\beq\label{Z}
 Z(g) =\sum_{G\in \C C_f} w(G).
\eeq
We define the corresponding probability measure on finite causal
triangulations by
\beq\label{R}
\rho_{CT}(G)={w(G)\over Z(g)}.
\eeq
The function 
$Z(g)$  can be computed \cite{Ambjorn:1998xu} by decomposing the sum over graphs into 
\bea Z(g)&=&\sum_{n=1}^\infty Z(g;n),\nn\\
Z(g;n)&=& \sum_{G\in {\C C}_{n+1}} g^{1+\Delta(G)}.\eea
$Z(g;n)$ is evaluated  by using \eqref{sigma}, and counting the graphs
in ${\cal C}_{n+1}$ by
building them up successively from the slices
$\{\Sigma_0,\ldots,\Sigma_{n-1}\}$.  The number of ways of connecting
$l_{k+1}$ vertices in $S_{k+1}$ with $l_{k}$ vertices in $S_{k}$ is
${l_k+l_{k+1}-1\choose l_k-1}$ and so, taking into account the marked edge,  
%
\bea
Z(g;n)&=&g\sum_{l_i\ge1,\atop n\ge i\ge 1}
\left( \prod_{k=1}^{n-1}
  {l_k+l_{k+1}-1\choose l_k-1}
  \right)
   g^{2(l_1+\ldots +l_{n})}.\eea 
Summing over $l_1$,  and using the binomial expansion of\\
   $(1-x)^{-l}$,
gives
\bea   Z(g;n)&=&g (\frac{X_1}{1-X_1})\nn\\&\times&\sum_{l_i\ge1,\atop n\ge i\ge 2}
\left( \prod_{k=2}^{n-1}
  {l_k+l_{k+1}-1\choose l_k-1}
  \right)X_2^{l_2}
   g^{2(l_3+\ldots +l_{n})} ,\eea
   where
   \beq X_{k+1}=\frac{g^2}{1-X_k},\quad X_1=g^2.\label{Xseq}\eeq
   Summing successively  over $\{l_2,\ldots\}$,    we find that
\bea
Z(g;n)&=&g\prod_{k=1}^{n} \frac{ X_k}{1-X_k}.\label{Zresult}\eea
%
The recursion \eqref{Xseq} is straightforward to solve and has the following properties:
\bea X_k&\uparrow& X^*=\frac{1-\sqrt{1-4g^2}}{2}\quad \mathrm{as~} k\uparrow\infty\mathrm{~for~}g<\half;  \nn\\
X_k&=&\half\frac{k}{k+1}\quad \mathrm{at~} g=\half.\label{Xprops}\eea
It follows that $Z(g)$ is analytic in the disk $\abs{g}<\half$
 and has a critical point at $g=\half$. 
 
 To understand the nature of the critical point it is instructive to
 compute the average girth, defined to be the length of the cycle at half height, of finite surfaces of a fixed height 
 \beq  L(n)=Z(g;2n)^{-1}\sum_{G\in {\C C}_{2n}} \abs{S_n}\,  g^{1+\Delta(G)}.\eeq
 Slightly more involved calculations than those above yield
 \beq L(n)<\frac{1}{\sqrt{1-4g^2}}.\eeq
Thus for any $n$ and $g<\half$ the average surface is like a long thin
tube closed off at the root end -- it is essentially
one-dimensional. However at  $g=\half$ we find that 
\beq L(n)=n+\quarter+\quarter\frac{1}{2n+1}.\eeq
%
which indicates that the  average surface at criticality is  two-dimensional.  To show that the Hausdorff dimension
$d_h$ defined in \eqref{Hausdorffdim} is indeed 2 we need to consider the tail distribution of large surfaces contributing to $Z(g)$. This
only makes sense at the critical point $g=\frac 12$ since only there does the mean area of
surfaces  diverge  due to the analyticity of
$Z(g)$ for $|g|<\frac 12$. The standard way to proceed is to condition the
distribution defining $Z(\frac 12)$ in \eqref{Z} on surfaces of fixed
finite area $N$ and take the limit  $ N\to\infty$ to get the
appropriate ensemble of infinite surfaces. We do this in the next
subsection by showing that the limit in question actually is equivalent
in a precise sense to the limit obtained in
Theorem~\ref{thm1}. Subsequently, in Section~\ref{sec:hausdorff},
we show that $d_h=2$ almost surely.


\subsection{Bijection between CT and planar trees}
\label{sec:3.2}
 We begin by showing that causal triangulations are in  one to
one correspondence with rooted planar trees.
 
\begin{figure}
\begin{center}
 \includegraphics[scale=0.5,angle=-90]{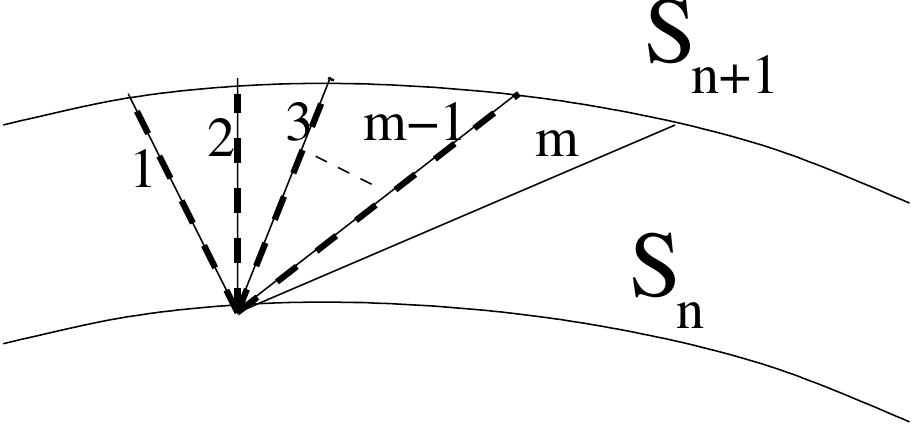}
 \end{center}
\caption{The bijection from $G\in{\C C}$ to $T\in{\C T}$: the dashed edges are assigned to $T$.}
\label{fig:1}       
\end{figure}

\begin{figure}
\begin{center}
  \includegraphics[scale=0.5]{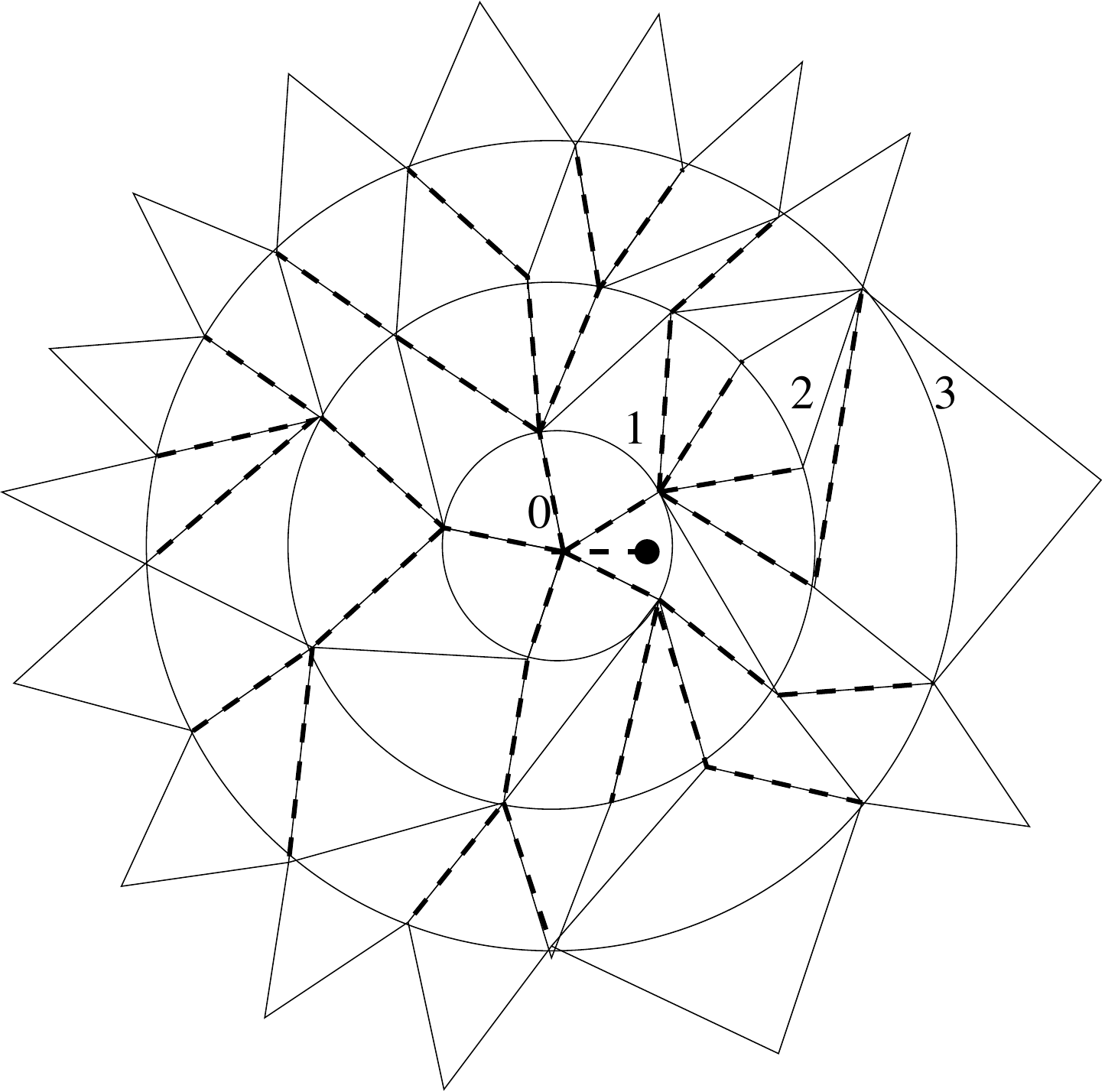}
  \end{center}
\caption{The bijection from $G\in{\C C}$ to $T\in{\C T}$: this example shows the tree equivalent to the triangulation of Fig.\ref{fig:0}. The dashed lines show the edges of the tree, including the new edge $(r,S_0)$.}
\label{fig:2}       
\end{figure}

Let $G\in{\C C} $.   We define a planar rooted tree $T=\beta(G)$
inductively w.r.t. height of edges in the following way:
\begin{enumerate}
\item The vertices of $T$ are those of $G$ whose  height is at most $h(G)-1$ 
  together with a new vertex $r$ which is the root of $T$ and whose
  only neighbour is $S_0$. 
\item All edges from $S_0$ to $S_1(G)$ belong to $T$ and the marked edge
  is the rightmost edge with respect to the edge $(r,S_0)$. 
\item For $n<h(G)-1$
 assign the edges emerging from a vertex $v\in S_n(G)$ and ending on
 $S_{n+1}(G)$ an ordered integer label  increasing by one each time in
 the  clockwise direction as shown in Fig.\ref{fig:1}. All edges
 except the edge with highest label belong to $T$ and have the same
 clockwise ordering.
\end{enumerate}
Fig.\ref{fig:2} shows an example of the application of these rules. 
Note that if the height of a vertex in $G$ is $n$ then its height in
$\beta(G)$ is $n+1$, i.e.  vertices in $S_n(G)$ are in $ D_{n+1}(T)$, $ n<h(G)-1$.

Conversely, let $T$ be a rooted planar tree. Then the inverse image
$G=\beta^{-1}(T)$ is obtained as follows: 
\begin{enumerate}
\item Mark the rightmost edge connecting $D_1(T)$ and $D_2(T)$. Delete the root of $T$ and the edge joining it to $D_1(T)$. The
  remaining vertices and edges of $T$  all belong to $G$ and $D_1(T)$
  becomes $S_{0}$, the root of $G$.
\item For $n\ < h(T)$ insert edges joining vertices in
  $D_{n+1}(T)$ in the circular order determined by the planarity of $T$; this creates the sub-graphs  $S_{n}(G)$.
\item For every vertex $v\in D_n(T)$, $ 2\leq n\leq h(T)-1$, that is not of
 order $1$ in $T$ draw an edge from $v$ to a vertex in $S_{n}(G)$
 such that the new edge is the most clockwise emerging from $v$ to
 $S_{n}(G)$ and does not cross any existing edges.
\item For every vertex $v\in D_n(T)$, $2\leq n\leq h(T)-1$, of order $1$ in $T$ draw
 an edge from $v$ to the unique vertex in $S_{n}(G)$ such that the
 new edge does not cross any existing edges. 
\item If $h(T)<\infty$, decorate the edges of the cycle of maximum height with triangles.
\end{enumerate}

A mapping equivalent to $\beta$ is described in \cite{Krikun}. For $G\in{\C C}_f$ these mappings are variants of Schaeffer's bijection
\cite{schaeffer:1998,schaeffer:2001}. Indeed, deleting the edges in $S_n(G)$ for all $n$ and identifying the vertices of maximal height $h(G)$ one
obtains a quadrangulation to which Schaeffer's bijection can be
applied; here the labelling of the vertices equals the height
function. As we have seen, the bijection extends in this case to
arbitrary infinite planar trees. For an extension to more general
planar quadrangulations see \cite{chassaing:2006}.  

This  construction shows that $\beta:{\cal C}\to {\cal T}$ is
a bijective map from $\tilde{\cal C}_N$, the set  of causal triangulations
of area $2N$, onto ${\cal T}_{N+1}$ and from $\cal C_\infty$ onto $\cal T_\infty$.
Moreover, defining the metric $d_{\cal C}$ on $\cal C$ by
\beq
d_{\cal C}(G,G') = \inf\{\frac{1}{R+1}\,:\; B_R(G)=B_R(G')\}\,,
\eeq
the map  is an isometry.

Now define the finite area probability distributions $\rho_N$ corresponding to \eqref{Z} and \eqref{R} at
$g=\frac 12$ by 
\beq\label{tildemuN}
\rho_N(G) = \tilde Z_N^{-1} 2^{-(1+\Delta(G))}\,,\quad G\in \tilde{\cal C}_N\,,
\eeq
where
\beq 
\tilde Z_N=\sum_{G\in\tilde\C C_N}  2^{-(1+\Delta(G))}.
\eeq

 The following result gives the relationship between generic random trees and infinite CTs.
\begin{theorem}\label{CDTequivGRT}
Let $\bar\mu_N$ and $\bar\mu$ be the measures defined by \eqref{muNdef} and
\eqref{mudef} corresponding to the generic, critical GW process with $p_n=2^{-(n+1)},\, n\geq
0$. Then 
\beq \label{equivN}
\rho_N(G) = \bar\mu_N(\beta(G))\,\quad G\in \tilde{\cal C}_N\,.
\eeq
The limit $\rho=\lim_{N\to\infty}\rho_N$ exists and is a probability
measure on ${\cal C}_\infty$ and is given by
\beq \label{equivinfty}
\rho(A) = \bar\mu(\beta(A))
\eeq
for any event $A\subseteq {\cal C}_\infty$.
\end{theorem}

\begin{proof}Existence of the limit and \eqref{equivinfty}
follow immediately from \eqref{equivN} and Theorem~\ref{thm1}. To
prove \eqref{equivN} consider a graph $G\in \C C_f$ and the 
corresponding tree $T=\beta(G)$. Every vertex  in $S_{i+1}(G)$ has exactly 
one edge of $T$ connecting it to $S_i(G)$ and therefore
\beq 
\abs {S_{i+1}(G)}=\sum_{v\in D_{i+1}(T)} (\sigma_v(T) -1)\,,\quad i=0,
\ldots ,h(G)-1 .\label{sumrule} 
\eeq
 Hence, from \eqref{area} we have
\beq 2^{-(1+\Delta(G))}= \frac 12 \prod_{i=1}^{h(G)-1}\, 2^{-2\abs{
    S_{i}(G)}} = \prod_{v\in T\setminus r} 2^{-\sigma_v}\,.
\eeq
Comparing this with \eqref{treemeasure} identity \eqref{equivN} follows.
\end{proof}



Note that $\rho_N$ as given by \eqref{tildemuN} is the uniform
distribution on $\tilde{\C C}_N$, that is 
\beq
\rho_N(G) = \frac{1}{\sharp\tilde{\C C_N}}\,,\quad G\in \tilde{\cal C}_N\,,
\eeq
where $\sharp\tilde{\C C_N}$ is the number of elements in $\tilde{\C
  C}_N$ (and is given by a Catalan number). For this reason the ensemble $({\C
  C},\rho)$ may appropriately be called the \emph{uniform infinite
  causal triangulation}.
According to Theorem~\ref{thm1} the measure $\rho$ is concentrated on the
subset $\beta^{-1}({\cal S})$ of triangulations corresponding to trees
with a single spine. 

A result analogous to Theorem \ref{CDTequivGRT} has been obtained for general planar triangulations in \cite{angel:2003aa}. Finally we observe that the present relationship
between trees and CTs is not the same as that introduced in \cite{DiFrancesco:1999em}; in that case
 the trees do not in general belong to a generic random tree ensemble.

\subsection{Reduced models}
\label{reduced}

We now define two simplified ensembles derived from the infinite CTs. These 
are useful in proving recurrence of the uniform infinite CT but also provide  models
which are interesting in their own right.

Let the set ${\C R}$  consist of all infinite graphs constructed  from the 
non-negative integers regarded as a graph so that $n$ has neighbours $n\pm 1$, 
except for 0 which only has 1 as a neighbour, and so that there are
$L_n$ edges connecting $n$ and $n+1$. Note that these graphs, an example of which is shown
in Fig.5,  have
multiple edges contrary to those considered above (they are called multi-graphs in the mathematical literature). 

The R ensemble is defined on ${\C R}$    by introducing a 
 mapping $\gamma:\C C_\infty\to\C R$ which acts on $G\in \C C_\infty$ by collapsing all the edges in 
$S_k,\, k\geq 1,$  and identifying all the vertices $v\in S_k$
so there is only one vertex at each height but  all the edges connecting 
$S_k$ and $S_{k+1}$ are retained.   The measure on $\C R$ is then inherited from 
that on $\C C_\infty$ so that  for  integers
$0\le k_1<\dots<k_m$ and positive integers $M_1,\dots, M_m$ 
\bea \lefteqn{\chi_\RR(\{G'\in \C R: L_{k_i}=M_i,\, i=1\ldots
  m\})}
  \nn\\&=&\rho(\{G\in \C C_\infty:\abs{S_{k_i}(G)}+\abs{S_{k_i+1}(G)}=M_i,\, i=1\ldots m\})\nn\\
&=&\bar\mu(\{T\in\C S:\abs{D_{k_i+1}(T)}+\abs{D_{k_i+2}(T)}=M_i,\, i=1\ldots m\}).\nn\\\label{RRequivGRT}\eea
%
%
%
%
%
%
A related ensemble ${\RRP}$ is obtained by defining $\gamma$ to retain only the  
edges connecting $S_k$ and $S_{k+1}$ that belong to the tree $\beta(G)$ in which case
the measure on $\C R$ is determined by 
\bea\lefteqn{ \chi_\RRP(\{G\in \C R:L_{k_i}=M_i,\, i=1\ldots m\})}\qquad\qquad\nn\\
&=&\bar\mu(\{T\in\C S:\abs{D_{k_i+2}}=M_i,\, i=1\ldots m\}).\label{RRPequivGRT}\eea

\begin{figure}
\begin{center}
  \includegraphics[scale=0.35,angle=-90]{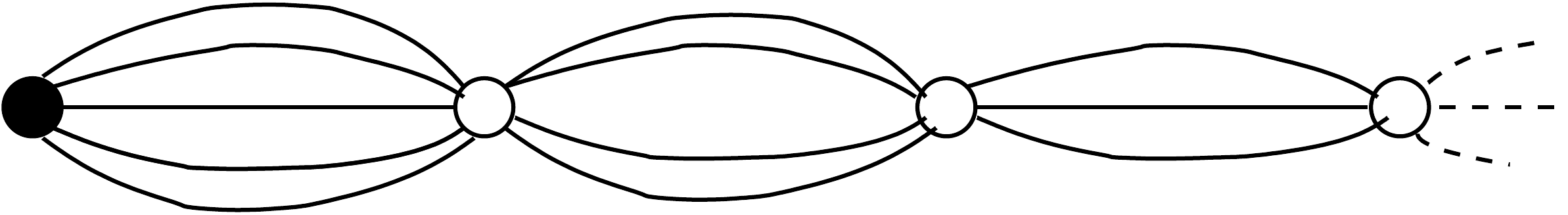}
  \end{center}
\caption{An example of $G\in{\C R }$.}
\label{fig:radial}       
\end{figure}

\section{Hausdorff dimension}
\label{sec:hausdorff}
As already indicated in the introduction the Hausdorff dimension of
a rooted infinite graph $G$ is defined by
\beq 
d_h = \lim_{R\to\infty}\frac{\log |B_R(G)|}{\log R}
\eeq
provided the limit exists. For the uniform infinite causal
triangulation we have the following result.
\begin{theorem}\label{hausdorffcausal} 
The Hausdorff dimension of a causal triangulation in ${\C C}_\infty$ is
$\rho$-almost surely equal to $2$. 
\end{theorem} 
Noting that 
\beq
|B_{R+1}(\beta(G))|\leq |B_R(G)| \leq 3|B_{R+1}(\beta(G))|\,,\quad
G\in{\C C_\infty}\,,
\eeq
this theorem is a direct consequence of Theorem~\ref{thm1} and the
following corresponding result for  generic random trees.

\begin{proposition}\label{hausdorfftrees}
For any generic random tree $({\C S},\mu_\infty)$ the Hausdorff dimension of $T\in{\C S}$ is
$\mu_\infty$-almost surely equal to $2$. 
\end{proposition} 

\paragraph{Proof}  
We actually prove a slightly stronger statement which is the
following: there exist positive constants $C_1$ and $C_2$ and for
$\mu_\infty$-almost all trees $T$ a constant $R_T >0$ such that
\beq\label{q1}
C_1(\log R)^{-2}R^2\leq |B_R(T )|\leq C_2 R^2\log R
\eeq
for all $R\geq R_T$.

We begin with the lower bound.  In \cite{Durhuus:2006vk}, Appendix 2, it is shown
that there are positive constants $c_0$ and $\lambda_0$ such that
\beq\label{q2}
\mu_\infty (\{T : \abs{B_R(T )}<\lambda R^2\} )\leq e^{-c_0\lambda^{-\half}}
\eeq
for $R>0$ and $0<\lambda <\lambda_0$.
Hence, for every $k>0$ we have 
\beq\label{q3}
\mu_\infty (\{T : \abs{B_R(T )}<k (\log R)^{-2} R^2\})\leq R^{-c_0k^{-\half}}
\eeq
if $R$ is sufficiently large.  Choosing $k\equiv C_1$ small enough it
follows that
\beq\label{q4}
\sum_{R=1}^\infty \mu_\infty (\{T : \abs{B_R(T )}<C_1 (\log R)^{-2}
R^2\})< \infty.
\eeq
By the Borel-Cantelli lemma we conclude that
\bea\label{q5}
\lefteqn{\mu_\infty (\{T : \abs{B_R(T )}<C_1 (\log R)^{-2} R^2\mathrm{~for~ infinitely~
many~}R \})}\qquad\qquad\qquad\qquad\qquad\qquad\qquad\qquad\nn\\ &=&0
\eea
and the lower bound follows.   

In order to establish the upper bound
we first prove that there exist constants $C_3$, $C_4>0$ such that
\beq\label{q6}
\mu_\infty (\{T : \abs{B_R(T )} >\lambda R^2\}) \leq C_3e^{-C_4\lambda}
\eeq
for all $\lambda$, $R>0$.  This is a slight generalization of Lemma
2.2 in \cite{Barlow:2005aa}.  Let $B_R^i$ denote the intersection of the ball of radius $R$, centred
at the spine vertex $s_i$, with the finite GW trees attached to $s_i$; then


\beq\label{q7}
\abs{B_R}\le R+ \sum_{i=1}^R\abs{ B_R^i}
\eeq
so it suffices to show that
\beq\label{q8}
\mu_\infty (\{T : \abs{B_R^1}+\ldots +\abs{B_R^R} >\lambda R^2\})\leq
C_3e^{-C_4\lambda}.
\eeq 
Since the $\abs{B_R^i}$ are independent and identically distributed random
variables  the Chebyshev inequality gives, for any $\theta >0$,
\bea\lefteqn{
\mu_\infty (\{T : \abs{B_R^1}+\ldots +\abs{B_R^R} >\lambda R^2\})}\qquad\qquad\qquad\nn\\
 &=&
\mu_\infty \left(\{T : e^{\theta (\abs{B_R^1}+\ldots +\abs{B_R^R})} > e^{\theta
\lambda R^2}\}\right)\nonumber\\
&\leq &e^{-\theta\lambda R^2}\expect{ \prod_{i=1}^R e^{\theta
\abs{B_R^i}} }{ \infty}   \nonumber\\
&=&e^{-\theta\lambda R^2}\left(\expect{  e^{\theta  
\abs{B_R^1}} }{ \infty }\right) ^R.\label{q10}
\eea
With the notation of \cite{Durhuus:2006vk}  we have
\beq\label{q11}
\expect{ e^{\theta \abs{B_R^i}}}{\infty} =g_R(e^\theta ),
\eeq
where $g_R(z)=f'(f_R(z))$ and
\beq\label{q12}
f_{K+1}(z)=zf(f_K(z)),~~ f_1(z)=z.
\eeq
Clearly \eqref{q12} defines $f_R$ inductively as an increasing analytic 
function
on $[0,1]$ such that $f_R(0)=0$ and $f_R(1)=1$.  By the
genericity condition, each $f_R$ is actually defined on a slightly
larger interval $[0,b_R]$ where $b_R>1$.  We will now show that we can
choose
\beq\label{q13}
b_R=1+{\beta\over (1+\alpha (R-1))^2}
\eeq
and for $z\in [1,b_R]$
we have
\beq\label{q14}
f_R(z)\leq 1+(1+\alpha (R-1))(z-1)
\eeq
for suitable constants $\alpha >1$ and $\beta >0$.
This will imply the bound \eqref{q8}.

We first choose $\rho_0>1$ such that $f(\rho_0)<\infty$.  Since
$f(1)=f'(1)=1$ there is a constant $k_1$ such that
\beq\label{q15}
f(z)\leq z+k_1(z-1)^2
\eeq
for $1\leq z\leq \rho_0$.
Setting $\beta\le \rho_0-1$ and
\beq\label{q16}  \alpha=(1+\beta)(1+k_1\beta)
\eeq
one can easily establish \eqref{q14} by elementary calculations and
induction.

Choosing $c>0$ sufficiently small we have
\beq\label{q16a}
f_R(e^{cR^{-2}})\leq e^{k_2/R}
\eeq
by \eqref{q14}, where $k_2>0$ is a constant.  Hence,
\beq\label{q17}
g_R(e^{cR^{-2}})\leq e^{k_3/R}
\eeq
for a suitable constant $k_3$.  Now taking $\theta =cR^{-2}$ in
\eqref{q10} we obtain the inequality \eqref{q8}.

The upper bound now follows in a similar way to the lower bound:  from
\eqref{q6} we have
\beq\label{q18}
\mu_\infty (\{T : \abs{B_R}>kR^2 \log R\} )\leq {C_3\over R^{C_4k}}.
\eeq
Choosing $k$ large enough we conclude that
\beq\label{q19}
\sum_{R=1}^\infty \mu_\infty (\{ T : \abs{B_R}>kR^2 \log R\} ) <\infty
\eeq
and the Borel-Cantelli lemma gives the upper bound for
$\mu_\infty$-almost every $T$.

\smallskip

We remark that it is a trivial consequence of this result that graphs
in the $\RR$ ensemble or the $\RRP$ ensemble likewise have Hausdorff dimension
$2$ almost surely. Moreover, defining the annealed Hausdorff dimension 
of a random graph $(\C G,\mu)$ by
\beq
d_h^{ann} = \lim_{R\to\infty}\frac{\log \expect{|B_R|}{\mu}}{\log R}\,,
\eeq
we have that $d_h^{ann}=2$ for any generic random tree as a consequence
of Lemma~\ref{Dinv}. It follows that this
holds for the uniform infinite CT and the $\RR$ and $\RRP$
ensembles as well.

\section{Recurrence of the uniform infinite causal triangulation}
\label{sec:recurrence}

 In this section we show  that random walk on graphs in the  uniform
 infinite CT and on graphs in the $\RR$ or $\RRP$  ensembles is almost surely
 recurrent. We start by giving a definition of recurrency.
  For a rooted graph $G$  let $\omega$  be a random walk on $G$ of length $n$ starting at the root at time 0 and let $\omega(t)$ denote the vertex
where $\omega$ is located after $t$ steps,  $ t \le n$.
 Simple random walk
  is defined in the standard
manner by attributing to $\omega$ the probability 
\beq 
p_\omega = \prod_{t=0}^{n -1}\sigma_{\omega(t)}^{-1}.
\eeq
The return probability is given by
\beq p_G(t)=\sum_{\omega:\omega(t)=r} p_\omega,\eeq
%
and the first return probability $p_G^0(t)$ by a similar sum restricted to walks which do not visit the root
at intermediate times,
 $\omega(t')\neq r$ for $0<t'<t$.
Note that $p_G(t)$ and $p_G^0(t)$ are defined for $t\leq n$ and are
otherwise independent of $n$. 
We say that random walk on $G$ is
recurrent if the random walk in the limit $n\to\infty$ returns to $r$
with probability $1$, that is if 
\beq 
\sum_{t=1}^\infty p_G^0(t) =1\,,\label{recurdef}
\eeq 
which is easily seen to be equivalent to (see \eqref{QP} below)
\beq 
 \sum_{t=1}^\infty p_G(t) =\infty\,.
\eeq
If $G$ is not recurrent it is called \emph{transient}.

There is a useful criterion for recurrency of an infinite
connected graph $G$  expressible in terms of the effective 
electrical resistance between the root and infinity when $G$ is considered
 as an electrical network in which each edge has resistance $1$. In
 the case of $\RR$ and $\RRP$ the resistance is
 straightforward to define as there is only one vertex at each height
 and it is simply
 \bea {R}_{G}\left(r,\infty\right)&=&\sum_{k=0}^{\infty} \frac{1}{L_k(G) }. \label{ResRR} \eea
 For $G\in \C C_\infty$ we define  $R_{G}\left(r,\partial
   B_K(G)\right)$ to be the resistance between the root and the vertex
 $v_{\mathrm{ top}}$ of the graph obtained from $B_K(G)$ by drawing in
 edges between all $v\in S_K(G)$ and a single new vertex
 $v_{\mathrm{ top}}$. We then obtain $R_G\left(r,\infty\right)$ by
 taking $K$ to infinity. The crucial result for our purpose is  
  \begin{theorem}
 Random walk on an infinite connected rooted graph is transient if and only
 if the effective resistance from the root to infinity is
 finite.\end{theorem} 
 \emph{Proof} The result is well known and a proof is given in e.g.  \cite{Lyons:2009}.

  For $\RR$ and $\RRP$  \eqref{ResRR} is sufficiently explicit  but for the
   infinite uniform CT we need an extra step. Define a cutset $\Pi$ in an
  infinite rooted graph $G$ to be a set
  of edges in $G$ such that a path from the root to infinity must
  include at least one member of $\Pi$. Denoting the number of edges in
  $\Pi$ by $|\Pi|$ we then have \cite{NW,Lyons:2009} 
 \begin{lemma}[Nash--Williams]\label{NashWilliams} If $\{\Pi_n\}$ is a sequence  of pairwise disjoint cutsets  in $G$ then
 \bea R_G\left(r,\infty\right)\ge\sum_n \abs{\Pi_n}^{-1}.\eea
 In particular, if the right hand side is infinite, then $G$ is recurrent.
 \end{lemma}
Choosing $\Pi_n$ to be those edges with one end in $S_n$ and one end in $S_{n+1}$ and applying the lemma
 gives the bound for $G\in \C C_\infty$
\bea R_{G}(r,\infty) &\ge&
\frac{1}{\abs{S_1}}+\sum_{k=1}^\infty\frac{1}{\abs{S_k} +
  \abs{S_{k+1}} }  \nn\\     &=& \sum_{k=0}^\infty
\frac{1}{\Delta(\Sigma_{k}) } \;.\label{ResCDT}\eea 

    Our proof of recurrence proceeds by establishing control over the
    right hand sides of \eqref{ResRR} and \eqref{ResCDT}. First we
    need                          
\begin{lemma}\label{tailprob} In the $\RRP$ ensemble   the probability that the number of edges in $G\in\C R'$ at height $n-1$  exceeds a fixed value $K$ is given by
\bea
\chi_\RRP(\{G:\abs{L_{n-1}}>K\})&=&\frac{K+n}{n}\left(1-\frac{1}{n}\right)^K,\; n>1,\label{PGRT}\eea
while for $\RR$  it is given by
\bea
\chi_\RR(\{G:\abs{L_{n-1}}>K\})&=&\frac{K+2n-1}{2n-1}\left(1-\frac{1}{2n}\right)^K,\; n>1.\nn\\\label{PGRT1}\eea
In the uniform infinite CT ensemble  the probability that for  $G\in \C C_\infty$ the number of triangles in
$\Sigma_{n-1}$ (equivalently the number of edges connecting $S_{n-1}$
to $S_n$) exceeds $K$ is given by 
\bea\rho(\{G:\Delta(\Sigma_{n-1})>K\})&=&\frac{K+2n-1}{2n-1}\left(1-\frac{1}{2n}\right)^K.\label{PCDT}\eea
\end{lemma}
\emph{Proof} These results are essentially well known. To prove \eqref{PGRT} note that from \eqref{RRPequivGRT}
\beq \chi_\RRP(\{G:\abs{L_{n-1}}>K\})=\bar\mu(\{T:\abs{D_{n}}>K\})\eeq
and the result then follows from Proposition 3.6 
 in \cite{ISI:000186418300011}. (Note that \eqref{PGRT} is the statement that $\abs{D_n}-1$ has the negative binomial distribution $\mathrm{NegBin}(2,1/n)$.) Using \eqref{RRequivGRT}, and noting
 that  $\Delta(\Sigma_{n-1})=\abs{S_{n-1}}+\abs{S_{n}}$, we see that
 \eqref{PCDT} and \eqref{PGRT1} are equivalent and, for $n\ge 2$,
 \beq \chi_\RR(\{G:\abs{L_{n-1}}>K\})=\bar\mu(\{T:\abs{D_{n}}+\abs{D_{n+1}}>K\}).\eeq
 Then using the proof of Proposition 3.6 
 in \cite{ISI:000186418300011}, \eqref{Xseq} and \eqref{Xprops},  we have
 \bea\lefteqn{ \bar\mu(\{T:\abs{D_{n}}=l_{n-1},\abs{D_{n+1}}=K-l_{n-1}\})}
 \nn\\
 &=&(K-l_{n-1})2^{-K+l_{n-1}-1}{K-1\choose
   l_{n-1}-1}\nn\\&&\qquad\qquad
\times\sum_{l_i\ge1,\atop n-2\ge i\ge 1}^\infty\left(\prod_{k=1}^{n-2}
   {l_k+l_{k+1}-1\choose l_k-1}\right) 4^{-(l_1+\ldots l_{n-1})},\nn\\ 
&=&(K-l_{n-1})2^{-(K-l_{n-1})-1}{K-1\choose l_{n-1}-1}(X_{n-1})^{l_{n-1}}\prod_{k=1}^{n-2} \frac{ X_k}{1-X_k},\nn\\
&=&\frac{(K-l_{n-1})2^{-(K-l_{n-1})}}{n(n-1)}{K-1\choose l_{n-1}-1}\left(\frac{n-1}{2n}\right)^{l_{n-1}},
\eea
and therefore
\bea\lefteqn{ \bar\mu(\{T:\abs{D_{n}}+\abs{D_{n+1}}=K\})}\qquad\qquad
\nn\\
&=&\sum_{l=1}^{K-1} \mu_\GRT(\{T:\abs{D_{n}}=l,\abs{D_{n+1}}=K-l\})\nn\\
&=&\frac{K-1}{(2n-1)^2}\left(1-\frac{1}{2n}\right)^K.\label{negbin}\eea
The lemma follows by summing over $K$. (Note that \eqref{negbin} is the statement that $\abs{D_n}+\abs{D_{n-1}}-2$ has the negative binomial distribution $\mathrm{NegBin}(2,1/(2n))$.)

\medskip
We can now establish the main result of this section:
\begin{theorem}\label{recurrence}
For a graph $G$ in the ensembles   $(\C R,\chi_\RR)$, $(\C
R,\chi_\RRP)$ or $(\C C_\infty,\rho)$ the effective resistance
between the root and infinity $R_{G}(r,\infty)$ is almost surely
infinite and random walk therefore almost surely recurrent. 
\end{theorem}
\paragraph{Proof} 
%
%
We give the proof in detail for the CT case and proceed by showing
that at large enough heights $n$ the number of triangles in slices
$\Sigma_{n-1}$ almost surely does not exceed the envelope function
$2an\log n$ where $a>1$ is 
a constant. 
First define the event that the number of triangles in $\Sigma_{n-1}$ exceeds the envelope
\bea \event{A}{a,n}&=&\{\Delta(\Sigma_{n-1})>2an\log n\},\quad n=1,2,\ldots.\label{event}\eea
Then from \eqref{PCDT} we find that
\beq \rho(\event{A}{a,n}) \le (1+2a\log n) n^{-a},\eeq
and so
\beq \sum_{n=1}^\infty \rho(\event{A}{a,n}) < \infty\,.\label{finitesum}\eeq
Hence, the Borel-Cantelli lemma can be applied to conclude that 
$\event{A}{a,n}$ occurs for at most finitely many $n$ with
probability $1$, that is for all graphs $G$ in a set of $\rho$-measure
$1$ there exists $N_G<\infty$ such that $\Delta(\Sigma_{n-1})\leq
2an\log n$ for all $n\geq N_G$. In particular, for such $G$ we have  
\bea \sum_{n=1}^\infty\frac{1}{\Delta(\Sigma_{n-1}) }\ge \sum_{n=N_G}^\infty\frac{1}{2an\log n }=\infty\, ,\eea
which, combining Lemma \ref{NashWilliams} with \eqref{ResRR} and
\eqref{ResCDT},   proves Theorem \ref{recurrence}  for $(\C
R,\chi_\RR)$ and $(\C C_\infty,\rho)$. To prove the theorem for
$(\C R,\chi_\RRP)$ we replace  \eqref{event} by  
\beq \event{A}{a,n}=\{{L_{n-1}}>an\log n\},\quad n=1,2,\ldots\eeq and proceed as above using \eqref{PGRT} in the next step.

\section{Spectral dimension of the R  and $\mathbf{R'}$ ensembles}
\label{sec:spectral}

We start  by  defining the generating functions \cite{Durhuus:2006vk}
\beq Q_G(x)= 1+\sum_{t=1}^\infty(1-x)^{\half t}
p_G(t)\eeq
and
\beq P_G(x)= \sum_{t=1}^\infty(1-x)^{\half t}
p^0_G(t).\eeq
%
%
The functions $Q_G(x)$ and $P_G(x)$ are related by the identity 
\beq\label{QP} Q_G(x)=\frac{1}{1-P_G(x)}.\label{PQrel}\eeq
In particular, it follows from \eqref{recurdef} that random walk on $G$ is recurrent if and
only if $Q_G(x)$ diverges for $x\to 0$.

By Theorem \ref{recurrence}  random walk is almost surely recurrent
for  the $\CDT$, $\RR$ and $\RRP$ ensembles. Assuming $Q_G$ has
asymptotic behaviour 
 \beq\label{spectraldim1}
 Q_G(x) \sim x^{-\alpha},\quad\alpha\in(0,1),\label{Qx}\eeq
for small $x$ then the return probability, $p_G(t)$, behaves asymptotically for large time as 
\beq\label{spectraldim2}
p_G(t)~\sim~t^{-\half d_s},
\eeq
where $d_s$ is the spectral dimension  of $G$ and is
related to $\alpha$ by a tauberian theorem
through
\beq\label{dsalpha}
d_s=2-2\alpha.
\eeq
 Note that if $d_s>2$ in \eqref{spectraldim2} then $Q_G(0)$ is finite and random walk
on $G$ is not recurrent.  In the borderline case 
$d_s=2$ we expect logarithmic corrections to the decay \eqref{spectraldim2} of $p_G(t)$  at large
$t$ and, if $G$ is recurrent, $ Q_G(x)$  to be logarithmically divergent at small $x$. 
We refer the reader to, for example,  \cite{flajolet} Sect. VI.3 and VI.11 for details on tauberian and transfer theorems.
Henceforth we shall take \eqref{dsalpha} as the definition of the
spectral dimension of $G$ where
\beq\label{alpha}
\alpha = \lim_{x\to 0}\frac{\log Q_G(x)}{|\log x|},
\eeq
which we assume exists.

The annealed spectral dimension $d_s^{ann}$ for a random graph is defined in the same
way as above by replacing $Q_G(x)$ in \eqref{alpha} by the ensemble average. 

\begin{theorem}\label{spectral}
For the ensembles   $(\C R,\chi_\RR)$ or $(\C R,\chi_\RRP)$  we have
that $d_s^{ann}=2$. Moreover, if $d_s$ exists almost surely then its
value is $2$ almost surely. 
\end{theorem}

\begin{figure}
\begin{center}
 \includegraphics[scale=0.5,angle=0]{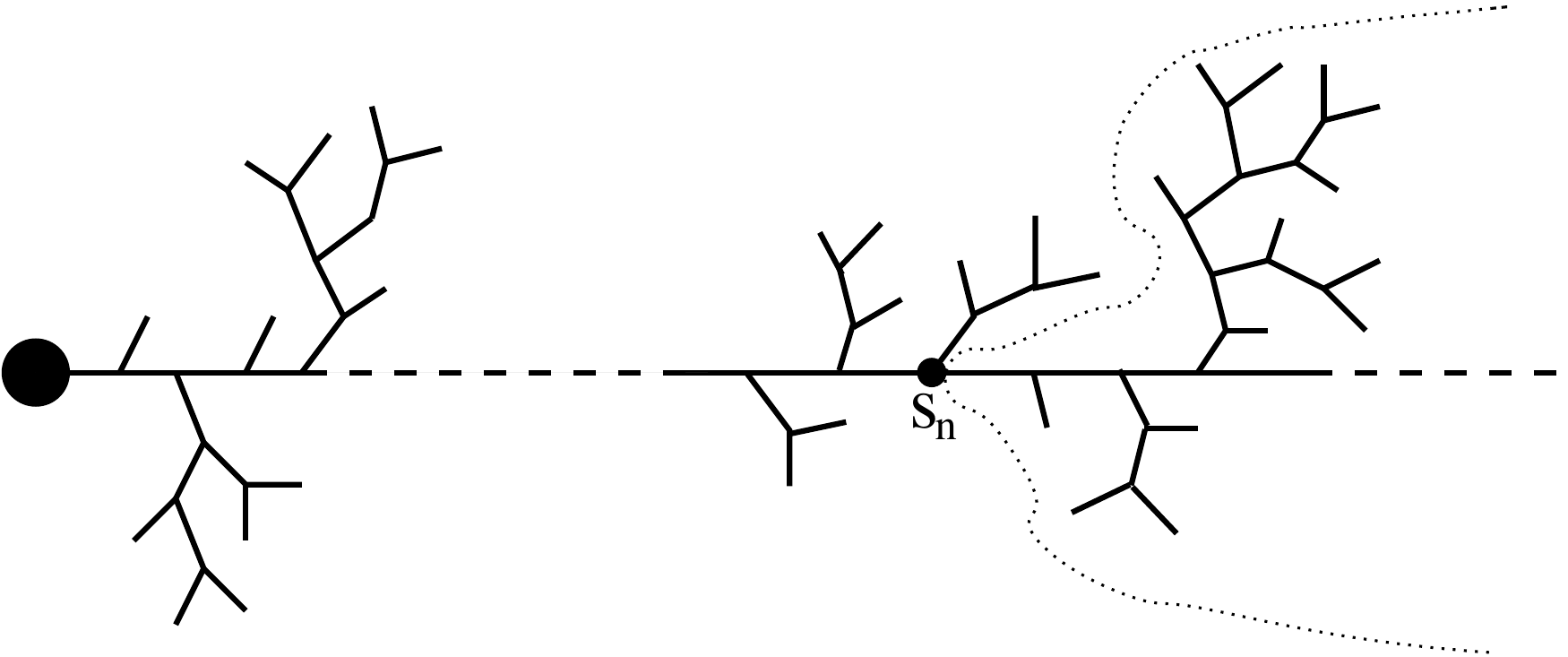}
 \end{center}
\caption{Example of a tree contributing (through  \eqref{RRequivGRT}) to \eqref{6.7} and \eqref{6.8}. Only the edges to the right of the dotted boundary line are included in the sum in \eqref{6.8}.}
\label{fig:Select}       
\end{figure}

We give the proof for $\RR$, that for $\RRP$ being essentially identical.
To prove the theorem we need
\begin{lemma}\label{Qlog} There is a constant $c>0$ such that
\beq \expect{Q_{G}(x)}{{\RR}}\le c \abs{\log x}\,.\eeq
\end{lemma}
\emph{Proof}.
Let $P_G(x;n)$ denote the generating function for first return to
vertex $n$ of a  random walk on a fixed graph $G\in\C R$ which leaves
$n$ in the direction of $+\infty$ with probability $1$ and let $Q_G(x;n)$ denote the
generating function for the corresponding return probabilities. The
generating function satisfies the recurrence relation  
\beq P_G(x; n-1)=\frac{(1-x) (1-u_G( { n}))}{1-u_G( { n})
P_G(x; n)},\label{6.1}\eeq
where 
\beq u_G( { n})=\frac{L_{n} ( { G})}{L_{n-1} ( { G})+L_{n} ( { G})}\eeq
is the probability that when the walk is at $n$ the next step is to $n+1$.
Defining $\eta_G(x; n)$ through
\beq P_{G}(x; n)=1-L_n( { G})^{-1}\eta_{G}(x; { n})\label{wk1}\eeq
and rearranging \eqref{6.1} gives
\bea\lefteqn{ \frac{1}{\eta_{G}(x; { n-1})}} \qquad\quad
 \nn\\
&=&\frac{1}{\eta_{G}(x;
  n)}+\frac{1}{L_{n-1} ( { G})} -\frac{ xL_{n-1} ( { G})}{\eta_{G}(x;
  { n})\eta_{G}(x; { n-1})}.\label{master}\eea
It follows that for $N\geq n$
\bea\label{recursion} \lefteqn{\frac{1}{\eta_{G}(x; { n-1})}}\nn\\
&=&\frac{1}{\eta_{G}(x; {
    N})}+\sum_{k=n-1}^{N-1}\frac{1}{L_{k} ( { G})} -x\sum_{k=n-1}^{N-1}\frac{
  L_{k} ( { G})}{\eta_{G}(x; { k})\eta_{G}(x; { k+1})}.\nn\\\eea
Note that since $P_G(x,k)<1$ we have $\eta_{G}(x;k)>0$ and
\eqref{recursion} implies that for $n\leq N$  
\beq \frac{1}{\eta_{G}(x; { n-1})}\le \frac{1}{\eta_{G}(x; { N})}+\sum_{k=n-1}^{N-1}\frac{1}{L_{k} ( { G})}.\eeq
Using \eqref{wk1} and \eqref{PQrel}, we then obtain
\bea Q_{G}(x; { n})&\le& L_{n} ( { G})\left( \frac{Q_{G}(x; { N}) }{ L_{N} ( { G}) }+\sum_{k=n}^{N-1}\frac{1}{L_{k} ( { G})} \right)\nn\\
&\le& L_{n} ( { G})\left( \frac{2 }{x L_{N} ( { G}) }+\sum_{k=n}^{N-1}\frac{1}{L_{k} ( { G})} \right),\label{6.7}\eea
where we have used the trivial bound $Q_{G}(x; { N})\le  2 x^{-1}$. We first maximise the
quantity in brackets in \eqref{6.7}  by including only
those edges inherited under $\gamma$ (c.f.  \eqref{RRequivGRT}) from the infinite tree whose root is  at $n$, as shown in Fig.\ref{fig:Select}, and which is distributed according to $\bar\mu$ as a consequence
of Theorem \ref{thm1}. Having done this the 
prefactor $L_{n} ( { G})$ is independent of the rest of the expression and taking expectation values gives
\bea \expect{Q_{G}(x;n)}{{\RR}}&\le& \expect{L_{n} ( { G})}{\RR}
\expect{ \frac{2 }{x \abs{D_{N-n+1} }}+\sum_{k=1}^{N-n}\frac{1}{\abs{D_{k} }   }     }{{\bar\mu}}\nn\\ 
&\le&  c'(n+2)\left(\frac{2}{x (N-n+1)}+\sum_{k=1}^{N-n}\frac{1}{k}\right),
\label{6.8}\eea
where we have used Lemma \ref{Dinv} and $c'$ is a constant.  Choosing $N=[x^{-1}]$ yields
\beq \expect{Q_{G}(x;n)}{{\RR}}\le c''(n+2)\,\abs{\log x}\eeq
and Lemma \ref{Qlog} follows by setting $n=0$.

\paragraph{Proof of Theorem \ref{spectral}} Theorem \ref{recurrence}
implies that $Q_G(x)$  diverges almost surely as $x\to 0$. Since
$Q_G(x)$ is a decreasing function of $x$ it follows that
$\expect{Q_G(x)}{\RR}$ and $\expect{Q_G(x)}{\RRP}$ diverge for $x\to 0$
and hence $d_s^{ann}\leq 2$. On the other hand, Lemma~\ref{Qlog}
implies $d_s^{ann}\geq 2$. 

It remains to show that $d_s\geq 2$ almost surely. We exploit the fact
that the logarithmic divergence of the ensemble average given by Lemma~\ref{Qlog} implies
that there cannot be a set  of non-zero measure of graphs whose $Q_G(x)$  diverges faster than logarithmically as $x\to 0$. 
For $0<x<1$ let
$${\cal A}_x = \{G\in {\cal R}\,:\; Q_G(x)> 1\}\,.$$  Since $Q_G(x)$
diverges almost surely and is decreasing in $x$ the sets ${\cal
  A}_x$ increase to a set of measure $1$ so that $\chi_\RR({\cal A}_x)\to
1$ for $x\to 0$. From Jensen's inequality we
get
\bea \lefteqn{\int_{{\cal A}_x} \log Q_G(x)d\chi_\RR}\qquad\qquad\nn\\
 &\leq& \chi_\RR({\cal
  A}_x)\log\left( (\chi_\RR({\cal A}_x))^{-1} \int_{{\cal
      A}_x}Q_G(x)d\chi_\RR\right)\nn\\ 
&\leq& \chi_\RR({\cal A}_x)\log\left((\chi_\RR({\cal A}_x))^{-1} \expect{Q_G(x)}{\RR}\right).
\eea
Dividing by $|\log x|$ and using Lemma~\ref{Qlog} then gives
\beq
\lim_{x\to 0} \expect{\max\{\frac{\log Q_G(x)}{|\log x|}, 0\}}{\RR}\; =\; 0\,.
\eeq
Assuming, as we do, that the limit \eqref{alpha} exists almost surely
this shows by the dominated convergence theorem that the limit $\alpha$
is non-positive, that is $d_s\geq 2$ almost surely.

\begin{remark} Simple random walk on a graph $G\in{\cal R}$ can equivalently be considered
as (non-simple) random walk on the non-negative integers with transition
probabilities $\alpha_n = \frac{L_n}{L_n+L_{n-1}}$ to go from $n$ to
  $n+1$ and $\beta_n=\frac{L_{n-1}}{L_n+L_{n-1}}$ to go from $n$ to
  $n-1$ for $n\geq 1$ and
  probability $\alpha_0=1$ to go from $0$ to $1$. For general
  $\alpha_n,\,\beta_n\geq 0$ with $\alpha_n+\beta_n=1$ such a process
  is called a \emph{birth and death process} and  is well
  known (see e.g. \cite{Karlin}) to be recurrent if and only if
\beq
\sum_{n=1}^\infty \frac{1}{L_n} =\infty \quad\mbox{where}\quad
L_n=\prod_{k=1}^n\frac{\alpha_k}{\beta_k}\,.
\eeq
Clearly, the proof of almost sure recurrence of the $\RR$ and $\RRP$
ensembles could have been based on this observation instead of the
Nash-Williams criterion. The estimate for the generating function
$Q_G(x)$ for return probabilities obtained 
in the proof of Lemma~\ref{Qlog} immediately generalises to arbitrary birth and
death processes in the form 
\beq 
Q(x)\leq 1+ \sum_{n=1}^{N-1}\frac{1}{L_n} + \frac{Q(x,N)}{L_N}\,,\quad
N\geq 1\,,
\eeq
where $Q(x)=Q(x;0)$ and $Q(x;N)$ denotes the generating function for return
probabilities for the random walk with transition probabilities
$\alpha'_n,\,\beta'_n$ given by $\alpha'_n = \alpha_{n+N}$. This in turn can be used to obtain an
estimate on the spectral dimension of the generalised random walk in terms of the decay rate of
$L_n^{-1}$ for large $n$. In particular, if 
\beq
L_n\;\sim\; n^\eta\,,\quad \eta < 1\,,
\eeq
then using the bound $Q_N(x)\leq \frac 2x$ and setting $N= [x^{-1}]$
one obtains 
\beq
Q_G(x)\;\leq \; c\,x^{\eta-1}
\eeq
for some constant $c>0$. This implies that the spectral dimension $d_s$ obeys
\beq
d_s\;\geq 2\eta\,.
\eeq
However,
this bound is generally not saturated as is seen from the example of the
simple random walk on the non-negative integers, where $\eta=0$ and
$d_s=1$. But in the limiting case $\eta=1$ we do get an optimal bound as
shown above.
\end{remark}

\section{Conclusions}
\label{sec:7}
We have shown that the spectral dimension of the uniform infinite causal
triangulation is bounded above by $2$ almost surely. This result is
compatible with the general result \cite{EJP2001-23} that random
planar graphs are almost surely recurrent if the degree of vertices is
bounded and certain uniformity assumptions are satisfied. However, the 
uniform infinite CT does not satisfy these conditions; for example, although
high degree vertices are relatively improbable, the vertex degree is not bounded. 
We have also shown that the Hausdorff dimension is exactly 2 almost surely so these graphs satisfy the bound
\beq d_s\le d_h\label{dsvdh}\eeq
 almost surely even though they do not necessarily obey the uniformity assumptions
of \cite{Coulhon:2000}.
The related $\RR$ and $\RRP$ reduced models have spectral and Hausdorff dimension
 exactly two almost surely and therefore saturate the bound \eqref{dsvdh}.

It is natural to conjecture that the spectral dimension of the uniform
infinite CT equals $2$ almost surely. The best lower
bound known to us derives from a comparison with the uniform infinite planar 
tree, which is known to have spectral dimension $4/3$
\cite{Durhuus:2006vk,Barlow:2005aa}. Indeed, deleting edges in a
graph decreases the Laplace operator associated with the graph and thus
decreases the spectral dimension. Hence, deleting the edges in a causal
triangulation $G$ that do not belong to the corresponding tree
$\beta(G)$ we get from Theorem~\ref{CDTequivGRT} that the spectral
dimension of the uniform infinite CT is at least $4/3$. By a similar
argument one can show that the spectral dimension of the $\RR$ ensemble provides an upper bound on that of the
uniform infinite CT.  
One possible strategy to prove the conjecture  would be to
gain better control of the error represented by this upper bound. 

It is worth noting that most of the results we have proved would go through if $\beta(G)$ were in 
\emph{any} generic random tree ensemble; only the proof of Lemma \ref{tailprob} uses the fact that 
we are dealing with the uniform infinite tree ensemble, but this is just a technicality. For example  \cite{DiFrancesco:1999em} considers an action generalized from \eqref{weight} to include a dimer-like contribution, controlled by a fugacity $a$, which mimics some features of a higher dimensional curvature term in the action. This model at its critical point maps to  an infinite random tree ensemble with the offspring probabilities
\bea p_0&=&g,\nn\\
p_n&=&a^{-2}g^{n+1},\eea
where $g=a(1+a)^{-1}$ (note that when $a=1$ and the dimers have no effect we just recover the uniform infinite random tree); our results therefore extend the observation of universal $a$-independent features
made in \cite{DiFrancesco:1999em}.
%
One can check that  any GW tree with off-spring probabilities $p_n$ corresponds via  $\beta$ to a CT model with  ultralocal action in which each vertex $v$ contributes the factor
\beq p_{\sigma_f(v)-1}\;g^{\sigma_v} \eeq
to the weight, where $\sigma_f(v)$ is the forward degree of $v\in G$. If the GW tree is critical the CT model is critical at $g=1$. 
So
%
%
%
 there is a whole universality class of surface models based on the generic random trees and all having $d_s\le 2$ and $d_h=2$. This would be even more interesting if it were to transpire that they all have $d_s=2$ exactly.

%

\begin{acknowledgements}
This work was supported by the EU Research Training Network grant MRTN-CT-2004-005616
and by UK Science and Technology Facilities Council grant ST/G000492/1.
\end{acknowledgements}


\end{document}